\documentclass[12pt,preprint]{aastex}
\usepackage{natbib}

\newcommand{\etal}{\textit{et al.}~}

\newcommand{\microns}{$\mu$m }
\newcommand{\msun}{M$_{\sun}$ }

\slugcomment{}

\shorttitle{}
\shortauthors{S. Gezari}

\begin{document}

\title{Adaptive Optics Near-Infrared Spectroscopy \\ of the Sgr A* Cluster}

\author{S. Gezari, A.M. Ghez\footnote{Also affiliated with UCLA's Institute 
of Geophysics and Planetary Physics.}, E.E. Becklin, 
J. Larkin, I.S. McLean, M. Morris}
\vspace{1cm}
\affil{Department of Physics and Astronomy \\ University of California,
    at Los Angeles, CA 90095-1562}
\email{suvi@astro.ucla.edu}

\begin{abstract}
We present K-band $\lambda$/$\Delta\lambda$ $\sim$ 2600 
spectroscopy of five stars 
(K $\sim$ 14 - 16 mag) within 0.''5 of Sgr A*, the radio source 
associated with the compact massive object suspected to be a
2.6 x 10$^{6}$ \msun black hole at the center of our Galaxy.  
High spatial resolution of $\sim$ 0.''09, and good strehl ratios of $\sim$ 0.2
achieved with adaptive optics on the 10-meter Keck telescope 
make it possible to measure moderate-resolution spectra of these 
stars individually for the first time.  
Two stars (S0-17 and S0-18) are identified as late-type stars by the
detection of CO bandhead absorption in their spectra.  Their absolute
K magnitudes and CO bandhead absorption strengths are consistent with early 
K giants.  Three stars (S0-1, S0-2, and S0-16), with r$_{proj}$ $<$ 0.0075 pc 
($\sim$ 0.''2) from Sgr A*, lack CO bandhead absorption, confirming
the results of earlier lower spectral and lower spatial resolution observations 
that the majority of the stars in the Sgr A* Cluster are
early-type stars.  The absolute K magnitudes of the early-type stars suggest
that they are late O - early B main sequence stars of ages $<$ 20 Myr.  
The presence of young stars in the Sgr A* Cluster, so close to the central 
supermassive black hole, poses the intriguing problem of how 
these stars could have formed, or could have been brought, within its 
strong tidal field. 
\end{abstract}

%\keywords{stars: individual (\eps Eridani, \bet Pictoris) -- planets: 
%extrasolar}

\section{Introduction} \label{intsec}
The Sgr A* Cluster is a concentration of faint stars 
(K $\sim$ 14 - 16 mag) 
observed within 0.''5 of Sgr A*, the compact radio source located at the
dynamical center of our Galaxy.  Stellar velocities in the central
25 arcsec$^{2}$ have demonstrated the existence of 2.6 x 10$^{6}$ \msun
of dark mass confined to a volume less than 10$^{-6}$ pc$^{3}$, most likely
in the form of a central supermassive black hole
(Genzel et al. 1997; Ghez et al. 1998).  The dynamics of the Sgr A* Cluster
stars are critical for further constraining the location and distribution of 
the central dark mass to smaller scales.  Ghez et al. (2000) used the proper 
motion accelerations of three Sgr A* Cluster stars 
(S0-1, S0-2, and S0-4) to localize the central dark mass to 
within 0.''05 $\pm$ 0.''04 of Sgr A*, and to 
increase the implied dark mass density by an order of magnitude 
compared to that achieved with the proper motion velocities. 
The dark matter distribution within the orbits of the Sgr A* Cluster stars can
be probed with even greater detail if the radial velocities of the stars
can be added to their proper motions, to determine their three-dimensional
orbits.
 
The Sgr A* Cluster stars can be fit by stars 
of two different spectral types given their absolute K magnitudes, assuming
 a distance of 8 kpc (Reid 1993) and 3 mag of 
K-band extinction to the Galactic Center: 1) early-type ($\sim$ B) dwarfs, or 
2) late-type ($\sim$ K) giants.  
If the Sgr A* Cluster stars  
are early-type main sequence stars, then they must be relatively young
($<$ 20 Myr).  This would amplify, by introducing young stars deeper into
the potential well, the problem already posed by the presence of the He I
stars (t$_{age}$ $\sim$ 5 Myr; e.g., Krabbe et al. 1995) of how and
where young stars that are currently in the proximity of a supermassive
black hole are formed (Morris 1993; Sanders 1992).

Due to the high density of stars in the central arcsecond,
both high spatial resolution and moderate spectral resolution K-band
spectroscopy are required to determine the nature and the radial velocities
of the individual Sgr A* Cluster stars.
Thus far no experiment has satisfied both criteria.
Genzel et al. (1997) used speckle spectroscopy to obtain high
spatial resolution but
low spectral resolution ($\lambda$/$\Delta\lambda$ $\sim$ 35) spectra of
four stars in the Sgr A* Cluster.  While they did not detect CO
bandhead absorption indicative of late type giants,
at this low spectral resolution
they were also unable to resolve any of the He I and Br$\gamma$ features
expected from early-type dwarfs. 
Other experiments, which achieved the necessary moderate spectral resolution 
($\lambda$/$\Delta\lambda$ $\sim$ 3000), but only
seeing limited spatial resolution ($\theta_{res}$ $\sim$ 0.''3 - 0.''6),
provided spectra of the Sgr A* Cluster stars together 
as an unresolved cluster (Eckart et al. 1999; Figer et al. 2000).  
These composite spectra demonstrated a lack of CO bandhead 
absorption from the northern part of the
Sgr A* Cluster, and a weak detection of CO bandhead absorption from the southern
part, 
constraining the presence of late-type stars in the cluster, and suggesting
that the majority of the Sgr A* Cluster stars are early-type main sequence
stars.

Wih NIRSPEC behind adaptive optics on the 10-meter Keck
telescope, one can achieve the spatial resolution and strehl
ratios needed to obtain moderate-resolution (R $\sim$ 2600) spectra of these
stars individually.
Thus, for the first time, we have in principle the capability of determining
the spectral types and the radial velocities of the
Sgr A* Cluster stars in the immediate vicinity of our Galaxy's central black
hole.  The data obtained improve considerably on previous measures of
the K-band spectra of a few of these stars, but the presumably weak
absorption lines expected for OB main sequence stars remain below our detection
threshold, perhaps confused in part by local gas emission in the same lines.
This paper describes this important, but difficult experiment in detail,
in anticipation of continued efforts to push this technique to the
requisite sensitivity level.

\section{Observations} \label{obssec}

Simultaneous K-band (2.0 - 2.4 \microns) spectra and images of stars in 
the Sgr A*  Cluster were obtained with the facility near-infrared 
spectrometer NIRSPEC (McLean et al. 1998, 2000) behind the adaptive
optics (AO) system
(Wizinowich et al. 2000)
on the W.M. Keck II 10-meter telescope during three nights 
of observations on 2000 June 20-22 (UT).  Near-diffraction-limited spatial
resolution was achieved with adaptive optics 
using the R = 13.2 mag natural guide star located 30'' NE from Sgr A*.  
The observed AO point spread function (PSF) was composed of a 
near-diffraction-limited core on top of a seeing halo of FWHM $\sim$ 0.''25 - 
0.''30. Figure 1 shows
the FWHM of the PSF's core and the fraction of energy it contained during the 
spectroscopic observations.  The measurements were made with a 0.''5 radius 
aperture on the bright star IRS 16C in the t$_{int}$ = 10 sec 
images of the central 4.4'' x 4.4'' taken with the slit viewing camera (SCAM),
plate scale = 0.0171''/pixel.  The nights of June 20 and 
21 had similar spectroscopic conditions, with the PSF containing a median
of $\sim$ 30\% of the total energy in a 0.''09 core.  The night of June 22
had somewhat worse conditions for spectroscopy, with the PSF containing a median of 
$\sim$ 25\% of the total energy in a 0.''1 core.

Spectra were obtained with SPEC, which contains a 1024$\times$1024
InSb array, using a 2-pixel-wide slit (3.96'' x 0.036'') in 
low-resolution mode, resulting in a 
moderate spectral resolution of $\lambda$/$\Delta\lambda$ $\sim$ 2600.  
Figure 2 shows the position of the slit 
on the Sgr A* Cluster for each night of observations.   On June 20 
the slit was centered on S0-16 \footnote{New sources named using the 
convention from Ghez et al. (1998).}, a new and variable source coincident
with the position of Sgr A* detected by Ghez et al.
(2002), 
and on June 21 and 22 the slit was centered on the Sgr A* Cluster stars S0-1 
and S0-2 
together.
Table 1 lists the five stars in the Sgr A* Cluster observed
 under the slit (S0-1, S0-2, S0-16, S0-17, and S0-18), their projected
distance from Sgr A*, K magnitude, date they were observed,
total integration time, and the 
signal-to-noise ratio per pixel of the 
extracted spectrum for each night of observations.  
The slit was aligned so that the 
bright star IRS 16NW (K = 10.1 mag) 
was in the slit for each t$_{int}$ = 300 sec exposure, in order for 
the exposures to be easily shifted and added together in post-processing of 
the data. 
Spectra were also obtained of Tau Sco (B0-V) and HD 203638 
(K0-III)
for spectral standards, and BS7105 (B8-V) and BS7194 (A2-III)
for telluric standards.

\section{Data Reduction and Analysis} \label{redsec}

The following standard data reduction routines were performed using 
JIP, an analysis package developed by James Larkin (Larkin 1996):  
1) Observations of a ``dark spot'' of sky relatively devoid of stars 180" East 
and 65" South of Sgr A* were subtracted from each spectrum to
remove the atmospheric and instrumental background, 2) the spectra were
spatially dewarped, 3) the spectra were wavelength 
calibrated using a spectral map constructed from neon arc lamps, and 4)
the spectra were
divided by spectra of telluric standard stars in order to correct 
for atmospheric absorption features in the spectra.  The telluric standard 
spectra, featureless
in the K-band except for Br$\gamma$ absorption, were continuum
corrected (i.e., divided by a blackbody of the same temperature),
normalized, and their Br$\gamma$ absorption feature removed by
linear interpolation.

Due to the large number of bad 
pixels in the spectrometer array, an additional bad pixel removal routine 
was performed
on the reduced exposures.  In order to account for fluctuations of pixel
fluxes between exposures due to varying seeing conditions and AO performance, 
and to take advantage
of dithering between exposures, bad pixels were identified
by looking at the ratio of
pixel fluxes in an exposure to the
median in the stack of exposures.  Pixels with flux ratios 
outside of 3 $\sigma$ for an exposure, were replaced by the mean flux 
ratio times the median flux of that pixel in the stack.

To ensure the best atmosphere and instrumental correction, spectra were 
reduced using all possible combinations of telluric standard 
stars and dark spot exposures obtained each night.
The best matched telluric standard star and dark spot exposure
was determined
by the signal-to-noise ratios of the final extracted spectra.  
The signal-to-noise of the continuum for each star was measured in the 
wavelength range from 2.18 to 2.28 \microns, where the spectra 
presumably lack 
stellar features.  This large
range of wavelengths was used in order for the 
signal-to-noise calculation to include noise from systematic 
errors due to poor telluric absorption and emission correction.

In order to maximize the signal-to-noise of the faint Sgr A*
Cluster stellar spectra, the reduced exposures were 
weighted by the peak
of IRS 16NW (integrated from 2.20 to 2.28 \microns) in each exposure before 
being summed together, so that the exposures with the highest strehl 
ratios contributed the most to the final averaged 2D 
spectrum.  The peak flux of IRS 16NW in each spectroscopic
exposure varied during the observing nights due to changes
in AO performance, seeing conditions, and centering
of the slit.
Figure 3 shows the final 2D spectra for each night, and figure 4 shows the
spatial profile of 
the Sgr A* Cluster stars and the astronomical sky background level in the 
final 2D spectra averaged 
from 2.20 to 2.28 \microns, in average counts per pixel.  
It is evident from the poor quality of the June 22 spectrum that, in 
addition to the total integration time, the 
signal-to-noise ratios of the final spectra are very sensitive to the 
strehl ratios of the individual exposures.

A 1D spectrum for each star was 
extracted from the final 2D spectrum from each night by carrying out a
weighted
average across $\sim$ 0.''2 in the spatial direction, centered 
on the peak of the star, by weights set by the PSF profile of IRS 16NW
as measured in the 2D spectrum. 
1D spectra of the astronomical sky background of unresolved stars and gas, 
henceforth referred to as the 'sky', in the region between IRS 16NW and S0-16 on June 20, and the region 
between IRS 16NW and
S0-2 on June 21 and June 22, were also extracted by averaging across the 
same width in the spatial direction.  The signal-to-noise ratios of the
spectra extracted from the June
22 spectrum are comparable to the signal-to-noise of the sky (S/N$_{pix}$ $\lesssim$ 10), 
so they were not used in this analysis. 
Figure 5 shows the 1D spectra extracted for the Sgr A* Cluster stars observed 
on June 20 and June 21, compared to the sky spectrum extracted each night.  
The spectral
features expected in early-type stars (He I and Br$\gamma$) and late-type
stars (CO bandheads) are labeled with tick marks. 
The feature at 2.316 \microns, seen in emission on June 20 and in
absorption on June 21 for all of the Sgr A* Cluster stars and the
sky background, is an atmosphere line that was not completely removed in the 
data reduction.  Figure 6 shows spectra obtained of two
bright stars in the central parsec which show the strong spectral
features in the K-band characteristic of early-type He I stars (IRS 16NW) 
and late-type supergiant stars (IRS 7).

\section{Results} \label{datsec}

\subsection{Late-type spectral features:  CO bandhead absorption} \label{cosec}

CO bandhead absorption is easily identified in late-type stars
because of its distinctive line shape and bandhead spacing.  
In order to identify CO bandheads in the low signal-to-noise spectra of the
Sgr A* Cluster stars,
and to measure their radial velocity shifts, the extracted spectrum of each 
star was cross-correlated with the spectrum of the observed K0-III 
spectral standard, HD 203638.
Figure 7 shows the cross-correlations for the wavelength range from 2.29 to 
2.37 \microns, which includes the $^{12}$CO (2,0), $^{12}$CO (3,1), 
and the $^{12}$CO (4,2)
absorption bandheads, for each of the Sgr A* Cluster stars observed (solid
line) in
comparison to the cross-correlation for
the sky on the same night of observations (dotted line).  The
K0-III standard spectrum and the Sgr A* Cluster spectra were normalized and 
the slope of the continuum in this wavelength 
range was removed before the cross-correlation.  Three of the Sgr A* Cluster
stars, S0-1, S0-2, and S0-16, show no significant cross-correlation peaks
with the K0-III standard spectrum above the sky background.  
S0-17 and S0-18, on the other hand, show 7 $\sigma$ and 5 $\sigma$ 
cross-correlation peaks 
respectively with the K0-III spectrum, where 1 $\sigma$ is
determined by the standard deviation of the cross-correlation for the
sky for that night of observations.  In order to not include the negative peak
in the sky cross-correlation attributed to the atmosphere line at 2.316 \microns,
only the portion of the sky 
cross-correlation with positive pixel shifts was used to determine 
1 $\sigma$.  It should
be noted that the cross-correlations for S0-17 and S0-18 peak at the 
same pixel shift.  This may point to the cross-correlation 
peaks as being the result of 
features in the spectra not intrinsic to the stars (i.e. features due to 
atmosphere lines, or systematic errors introduced in the data reduction
process).  However, the other Sgr A* Cluster stars   
(S0-1, S0-2, and S0-16) do not show significant correlation 
peaks at this pixel shift, making it hard to attribute the 
cross-correlation peaks for S0-17 and S0-18 to systematic features in 
the spectra.  Assuming that the CO features are real, we derived a radial
velocity of -66 $\pm$ 17 km/s for S0-17 and -76 $\pm$ 13 km/s for S0-18.
The velocity shifts were measured by the peak of the gaussian fit to the 
cross-correlation
peak, and the error was determined by the jack-knife resampling method
(Babu \& Feigelson 1996), in which random half sets of data were cross-correlated
with the K0-III spectrum, and the dispersion of these velocity fits from the
original velocity shift is the 1 $\sigma$ error.

In addition to cross-correlation as a method for identifying CO
bandhead absorption, absorption strengths for the three strongest
bandheads in this wavelength range, $^{12}$CO (2,0),
$^{12}$CO (3,1), and $^{12}$CO (4,2),
were measured for the stars that showed cross-correlation peaks with the K0-III
standard (S0-17 and S0-18), at the appropriate wavelengths determined
by the velocity shift of
the peak of the cross-correlation.  Absorption strengths were
also measured for the sky at those wavelengths to estimate upper limits
on absorption strengths due to systematic errors.
Absorption strengths for the CO bandheads, 
[(1 - F$_{CO}$/F$_{cont}$) x 100], were measured using a bandwidth of 
$\Delta\lambda$ = 0.0055 \microns for the CO absorption and
continuum fluxes, where the continuum was measured to the 
left of the first CO bandhead at 2.29 \microns, 
in order to match the methods used in the Kleinmann and Hall (1986) 2.0 - 
2.5 \microns spectral
atlas of late-type standard stars.  The spectra were divided by the
slope of the 
continuum from 2.00 to 2.29 \microns before measuring 
absorption strengths, so that the slopes of the spectra would be comparable 
to the flat spectra in Kleinmann and Hall (1986) which were
directly ratioed by a telluric reference star that was not corrected for
its continuum.  The statistical error in the absorption strength
measurements was estimated by the jack-knife
resampling method.
Figure 8 shows the CO bandhead absorption strengths with 1 $\sigma$ error bars
for S0-17 and S0-18, in comparison to the sky for that night of observations,
the observed K0-III standard HD 203638 (plotted with diamonds), and
the K0-III standard {\it i} Cep from Kleinmann and Hall (1986) (plotted with
triangles).
Table 2 lists the absorption strengths measured for S0-17 and S0-18.

S0-17 and S0-18 both
have significant $^{12}$CO (2,0),
$^{12}$CO (3,1), and $^{12}$CO (4,2) absorption strengths consistent
with a K0 giant (within 3 $\sigma$) as measured for both the 
Kleinmann and Hall (1986) K0-III standard, and the K0-III standard 
observed in this analysis.  
The CO absorption cannot be attributed to systematic errors, since
the sky background demonstrates very little or no absorption at these same
wavelengths. 
The absorption strength measurements, in combination with the
cross-correlation peaks with the K0-III giant spectrum, are suggestive evidence that
S0-17 and S0-18 are in fact late-type giants.  The other sources, S0-1, 
S0-2, and S0-16, have higher signal-to-noise spectra than S0-17 and S0-18,
and are uncorrelated with the 
K0-III spectrum, strongly suggesting that these stars lack CO bandhead 
absorption, and therefore do not appear to be late-type stars. 

\subsection{Early-type spectral features:  He I and $Br\gamma$} \label{hesec}

In addition to a lack of CO bandhead absorption, early-type main sequence
stars can be identified by their He I absorption lines at 2.058 \microns and 
2.113 \microns, and H I Br$\gamma$ absorption line at 2.166 \microns.  
In the Hanson et al. (1996) 2 \microns spectral atlas of hot, luminous stars,
the strongest feature in the K-band spectra of late O - early B 
main sequence stars is the H I Br$\gamma$
 absorption line.  This also appears to be the line most sensitive
to stellar temperature, as the Br$\gamma$ absorption
 equivalent width monotonically increases from 1 $\AA$ for an O5 star
up to 7 $\AA$ for a B5 star.  The He I lines at 2.058 \microns  and 
2.113 \microns, on the other hand,
 have weak absorption equivalent widths ($\lesssim$ 1 $\AA$) in this 
range of spectral types, and
their strengths do not have an obvious dependence on temperature.  
In the spectral class range
we are interested in for the Sgr A* Cluster, the Br$\gamma$ line appears 
to be the most useful diagnostic for spectral classification.
Unfortunately, the Sgr A* Cluster spectra are heavily contaminated 
at 2.166 \microns by Br$\gamma$ emission
from ionized gas in the region.  This is demonstrated
in the 2D spectrum on June 20 and June 21 (Figure 4), where Br$\gamma$ is
seen in emission across the spectrum. At this spectral resolution, we 
cannot separate out any Br$\gamma$ absorption intrinsic to the Sgr A*
Cluster stars that may be
veiled by the strong background gas emission.  We can, however,
estimate the strength of the stellar Br$\gamma$ line by first
subtracting out the flux of Br $\gamma$ emission 
from the sky background, 
and then measuring the residual flux in the stellar spectrum.
The Br$\gamma$ line equivalent widths were measured at 2.166 \microns using a 
bandwidth of $\Delta\lambda$ = 0.02 \microns,
in order to allow for doppler shifts of the lines from the rest wavelength of
up to $\pm$ 1500 km/s.
Table 2 lists the resulting Br$\gamma$ line equivalent widths measured for the
Sgr A* Cluster stars after the line emission
from the sky has been subtracted out, with 1 $\sigma$ error bars determined
by the jack-knife resampling method.  

Within 2 $\sigma$, S0-17 and S0-18 do not demonstrate Br$\gamma$ absorption
in their spectra, consistent with their classification as early
K giants.  A suggestion of residual Br$\gamma$ emission
can be attributed to the incomplete removal of contributions from the background gas 
emission.  
The spectra of stars lacking CO absorption, S0-1, S0-2, and S0-16, have Br$\gamma$
equivalent widths that are within the range measured 
for late O - early B main sequence stars (within 2 $\sigma$).  However, 
given the large errors 
in the equivalent widths measured for the Sgr A* Cluster stars, 1 $\sigma$ 
$\sim$ 2$\AA$, and the 
 uncertainties in subtraction of the background gas contamination, the 
Br$\gamma$ measurements are not statistically significant and therefore are
not conclusive enough to classify or rule out that these stars are early-type.
In addition, difficulties
in measuring lines in this
 region of the spectrum arise from uncertainties in the 
removal of the Br$\gamma$ absorption line in the telluric standard spectrum.
Hanson et al. (1996) emphasize that this is their largest source of systematic
error in their Br$\gamma$ line equivalent widths.  

The He I 2.058 \microns line is also contaminated by emission from
background ionized He I gas.  However, even without the difficulties
introduced by background gas emission, both He I lines at 2.058 \microns and
2.113 \microns have absorption strengths in normal stars that are intrinsically too low
to be detected in the Sgr A* Cluster spectra, which have equivalent width
errors up to 3 $\AA$.
The very low signal-to-noise
ratios of the Sgr A* Cluster spectra, and the strong contamination by
background gas emission makes identification
of early-type spectral features intrinsic to the stars in the spectra
infeasible at this time.

\section{Discussion \& Conclusions} \label{dissec}

The velocities of the Sgr A* Cluster stars can be used to determine an upper
limit to their current distances from the central black hole by assuming 
that the stars are in a bound orbit, and thus that r $<$ 2GM$_{bh}$/v$^{2}$.
Combining the radial velocities for S0-17 and S0-18 reported in Table 2,
with their proper motions (Ghez et al. 2002), we derive three-dimensional
velocities of 730 and 320 km/sec, respectively, which bounds their positions
to within 0.04 and 0.22 pc.  This makes them both likely members of the 
Sgr A* Cluster, which has a projected radius of 0.02 pc. 

The three dimensional velocities 
measured for S0-17 and S0-18 cannot be used to 
better constrain the density of the enclosed
central dark mass, because their radial distances are greater than the smallest
radius of 0.015 pc measured by the proper motion experiment of Ghez et al. (1998).
However, if the radial velocities of the three stars identified as early-type
(S0-1, S0-2, S0-16) could be
measured with future higher spectral resolution spectroscopy sensitive to the
He I and Br$\gamma$ absorption lines expected to be present in these
stars, then their 3D velocities could constrain the central
dark mass density down to radii $<$ 0.0075 pc, and further establish the 
configuration of the central dark matter distribution
as a central supermassive black hole, as opposed to a cluster of stellar
remnants, or other more exotic forms of dark matter. 

The lack of CO bandhead absorption
in the moderate resolution
spectra of S0-1, S0-2, and S0-16 is direct evidence that these stars are 
not late-type stars, and with their absolute K magnitudes, that they may be 
late O - early B main sequence stars with ages less than 20 Myr.  The existence
of young stars in the immediate vicinity of a central supermassive
black hole is surprising, considering the extreme conditions expected to 
inhibit star formation in such an environment. 
Morris (1993) discusses how the strong
magnetic fields ($\sim$ mG), large turbulent velocities (v $\sim$ 10 km/s),
high temperatures, and the large tidal forces induced
by the central black hole in the Galactic Center would only enable star 
formation in gas clouds of very high densities.  Gas near the
Galactic Center would have to be compressed to densities 5 orders of 
magnitude higher than the densities currently inferred for nearby gas 
in order to avoid tidal disruption
and be able to gravitationally collapse to form stars.  Such a large 
density might be accomplished by a violent 
compression
of gas clouds caused by cloud collisions, 
stellar winds, supernova shocks, or the violent release of
accretion energy by the black hole.  Alternatively, the presence
of early-type 
stars in the Sgr A* Cluster might be explained by dynamical friction
acting on a massive young star cluster which formed at a substantial 
distance from the black hole and which migrates into the central parsec on
a time scale less than the lifetime of these relatively massive stars 
(Gerhard 2000, Kim \& Morris 2002).

K-band spectra of five stars in the Sgr A* Cluster demonstrate that
this concentration of faint stars mainly contains young, late O - 
early B main 
sequence stars, consistent with the low spectral resolution
results of Genzel et al. (1997) and the low spatial resolution
results of Eckart et al. (1999) and Figer et al. (2000),
along with a small number of older K giants.
The close proximity of the young stars in the Sgr A* Cluster 
to the 2.6 x 10$^{6}$ \msun black hole, is an interesting
challenge for either star formation theory or dynamical theory.  In either 
case, more can be 
learned about the star formation history of the 
Galactic Center, and about the detailed dark mass distribution there,
if the spectral types and radial velocities of the Sgr A* Cluster 
stars can be unambiguously determined.  This may be possible with future, 
higher spectral resolution adaptive optics spectroscopy.

\acknowledgments
\textit{This work has been supported by the National Science
Foundation Science (NSF) and Technology Center for Adaptive Optics, managed
by the University of California at Santa Cruz under cooperative agreement
No. AST-9876783, by individual NSF grant No. AST-9988397, and by the
Packard Foundation.}

\begin{deluxetable}{lcclcc}
\tabletypesize{\scriptsize}
\tablecaption{Log of Spectroscopic Observations \label{logtbl}}
\tablewidth{0pt}
\tablehead{
\colhead{Source} &
\colhead{r ('')\tablenotemark{a}} &
\colhead{K (mag)} &
\colhead{Date} &
\colhead{Total t$_{int}$ (min)} &
\colhead{S/N$_{pix}$}
}
\startdata
S0-1 & 0.11 & 14.9 & 21 June & 100 & 21 \\
  &  &   & 22 June & 80 & 9 \\
S0-2 & 0.15 & 14.1 & 21 June & 100 & 25 \\ 
  &   &   & 22 June & 80 & 10 \\
S0-16 & 0.046 & 15.3 & 20 June & 45 & 18 \\
S0-17 & 0.21 & 16.0 & 20 June & 45 & 17 \\
S0-18 & 0.44 & 15.1 & 21 June & 100 & 15 \\
   &  &    & 22 June & 80 & 9 \\
Sky & 0.37 & - & 20 June & 45 & 12 \\
    & 0.44 & - & 21 June & 100 & 12 \\
    & 0.30 & - & 22 June & 80 & 9 \\
\enddata
\tablenotetext{a}{Projected distance from Sgr A*.}
\end{deluxetable}

\begin{deluxetable}{lcccccc}
\tabletypesize{\scriptsize}
\tablecaption{Summary of Spectral Data \label{latetbl}}
\tablewidth{0pt}
\tablehead{
\colhead{Star} &
\colhead{Br$\gamma$ {\it{EW}}\tablenotemark{a}} &
\colhead{$^{12}$CO(2,0) {\it{AS}}} &
\colhead{$^{12}$CO(3,1) {\it{AS}}} &
\colhead{$^{12}$CO(4,2) {\it{AS}}} &
\colhead{Spectral\tablenotemark{b}} &
\colhead{Radial Velocity} \\
\colhead{ } &
\colhead{($\AA$)} &
\colhead{[(1-F$_{CO}$/F$_{cont}$)x100]} &
\colhead{[(1-F$_{CO}$/F$_{cont}$)x100]} &
\colhead{[(1-F$_{CO}$/F$_{cont}$)x100]} &
\colhead{Type} &
\colhead{(km/s)}
}
\startdata
S0-1 & 0.64 $\pm$ 1.22 & - & - & - & early & - \\ 
S0-2 & 1.27 $\pm$ 2.04 & - & - & - & early & - \\
S0-16 & 1.03 $\pm$ 1.91 & - & - & - & early & - \\
S0-17 & -3.60 $\pm$ 2.25 & 11.35 $\pm$ 2.50 & 14.76 $\pm$ 1.35 & 10.14 $\pm$ 1.96 & late & -66 $\pm$ 17 \\
S0-18 & -1.15 $\pm$ 1.67 & 13.53 $\pm$ 2.48 & 10.00 $\pm$ 1.80 & 8.93 $\pm$ 3.01 & late & -76 $\pm$ 13 \\
\enddata
\tablenotetext{a}{Equivalent widths measured after flux from background gas
emission was subtracted out.  Positive equivalent widths indicate 
absorption, negative equivalent widths indicate emission.}
\tablenotetext{b}{Spectral type determined by the presence (late-type) or 
lack (early-type) of CO bandhead absorption.}
\end{deluxetable}

\begin{figure}[tbp] \label{statsfig} \figurenum{1}
\begin{center}
\plottwo{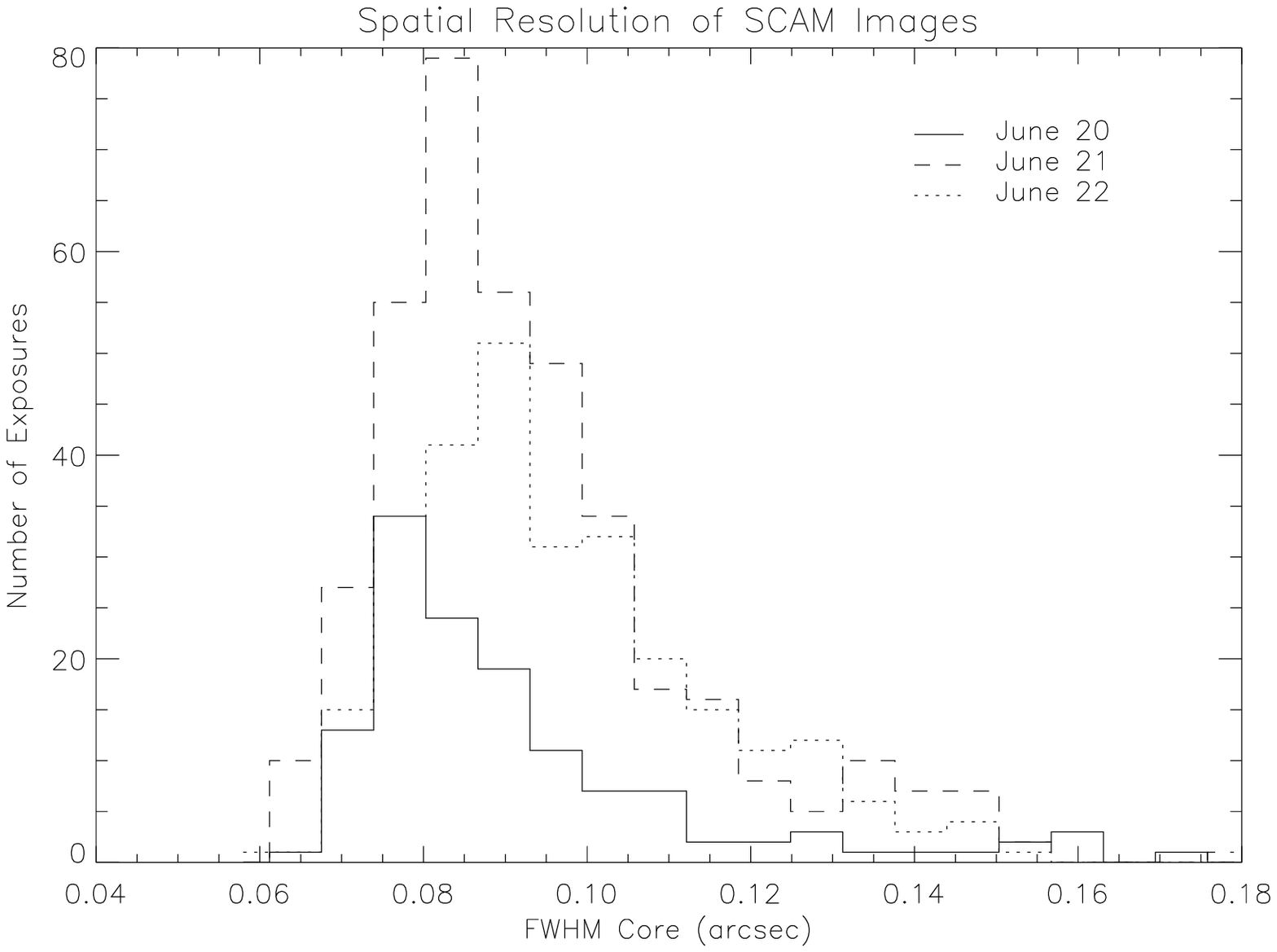}{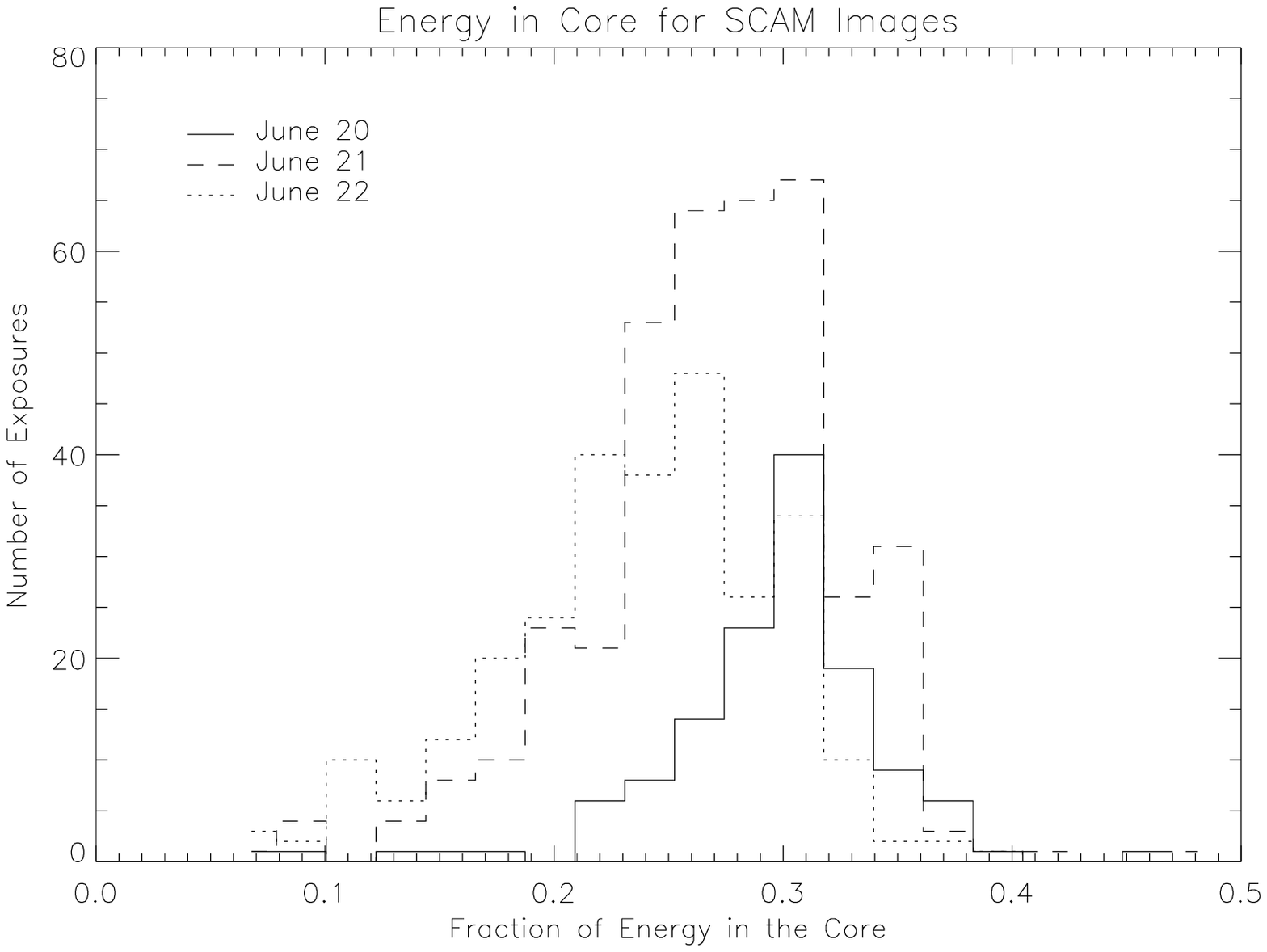}
\caption{Histograms of the FWHM (left), and fraction of energy in the core (right), of the AO corrected PSF in the t$_{int}$ = 10 sec SCAM 
images taken during each night of observations.}
\end{center}
\end{figure}

\begin{figure}[tbp] \label{slitfig} \figurenum{2}
\begin{center}
\plotone{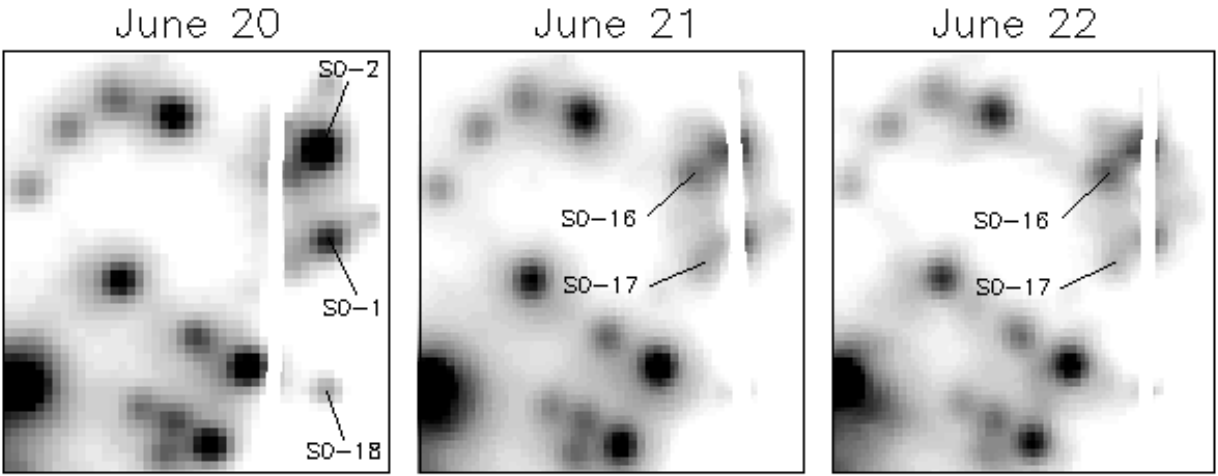}
\caption{Co-added SCAM images of the Sgr A* Cluster, clipped down to 
the central 0.77'' x 0.86'', to show the position of the slit
during the spectroscopic observations on June 20 (t$_{int}$ = 22 min),
June 21 (t$_{int}$ = 64 min), and June 22 (t$_{int}$ = 46 min).  Images are
oriented such that north is up and east is to the left.}
\end{center}
\end{figure}

\begin{figure}[tbp] \label{specfig} \figurenum{3}
\begin{center}
\plotone{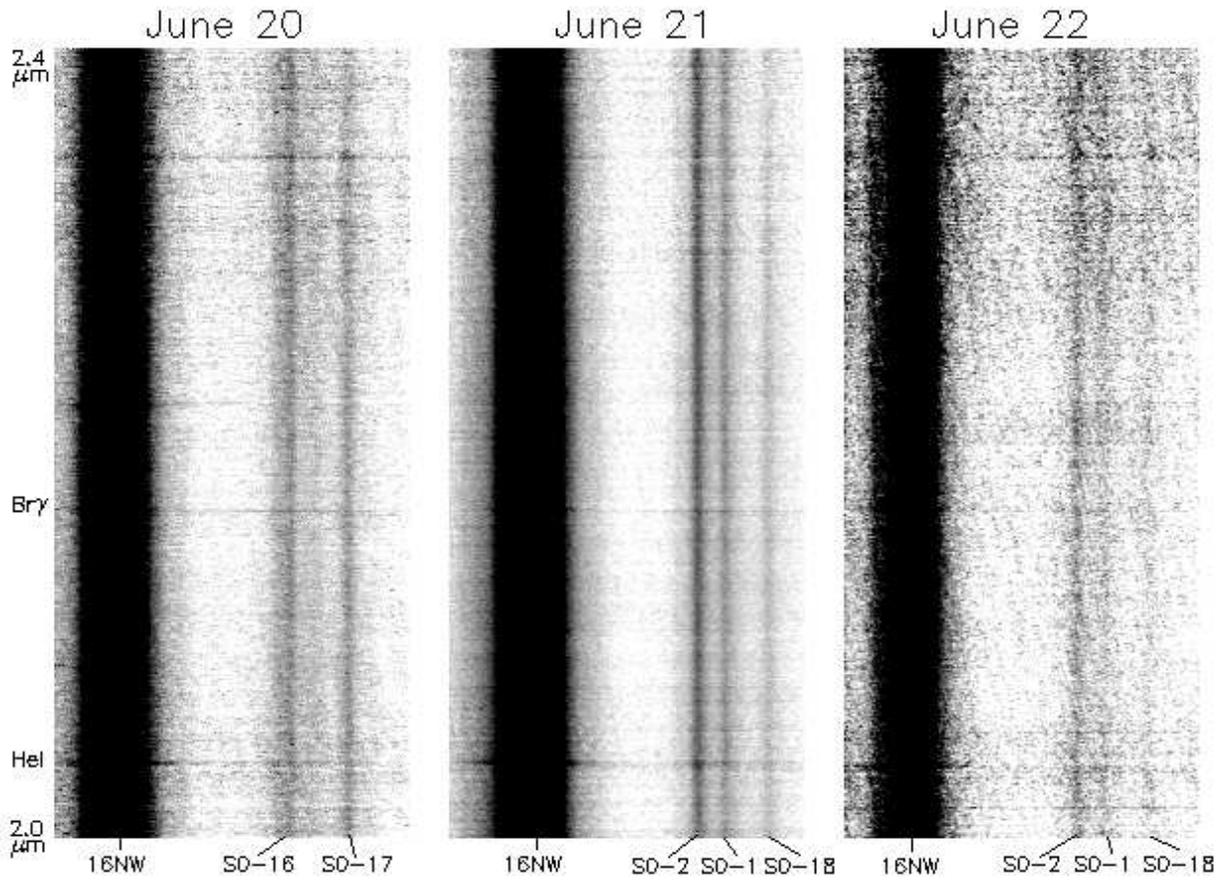}
\caption{Final 2D K-band spectra for each night of observations.}
\end{center}
\end{figure}

\begin{figure}[tbp] \label{peakfig} \figurenum{4}
\begin{center}
\plotone{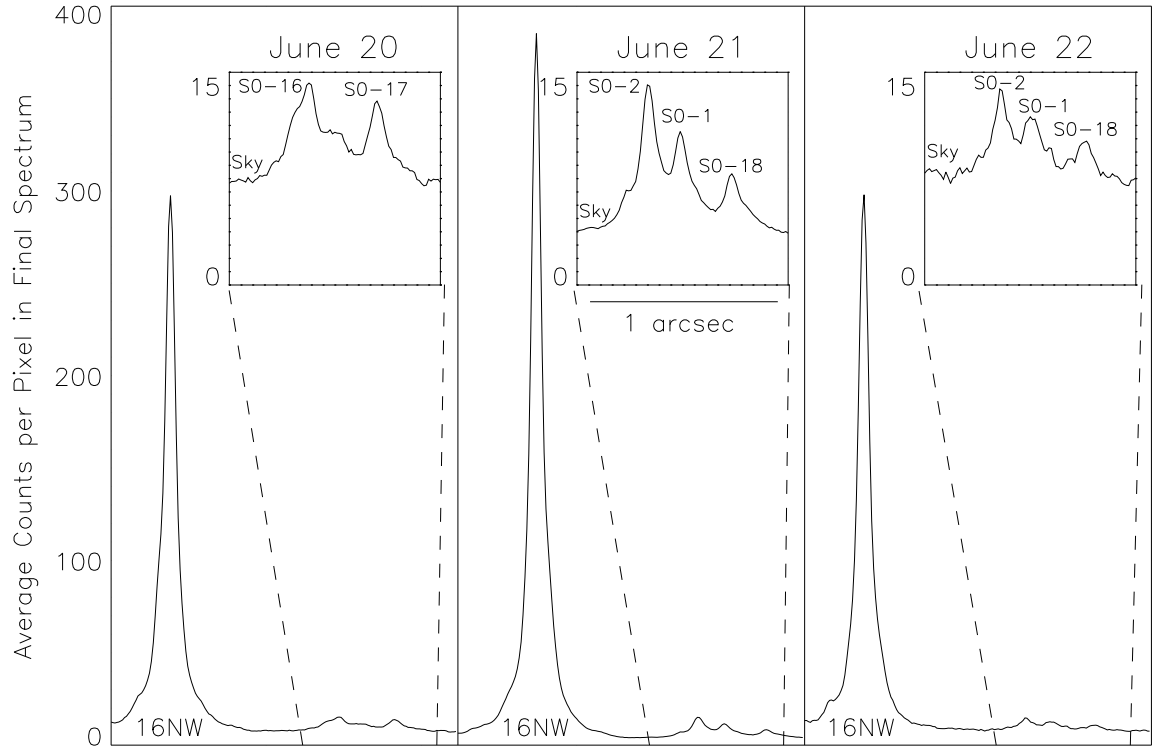}
\caption{Spatial profile of the final 2D spectra averaged from 2.20 to
2.28 \microns in counts per pixel.}
\end{center}
\end{figure}

\begin{figure}[tbp] \label{specsfig} \figurenum{5}
\begin{center}
\plotone{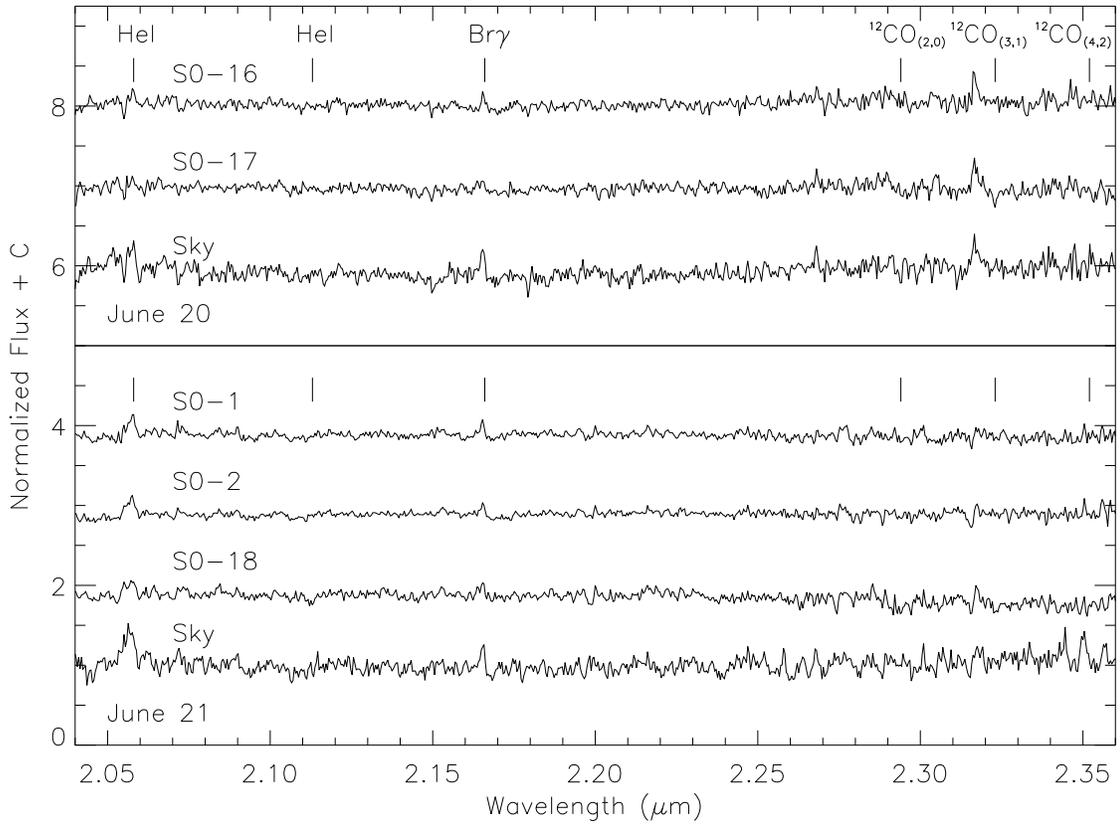}
\caption{K-band spectra of the Sgr A* Cluster stars observed
June 20 and June 21 in comparison to the sky for that night.  The spectra 
are normalized to their flux at 2.20 \microns and shifted up by 
an increment of 1, 2, 3, 5, 6, and 7 for S0-18, S0-2, S0-1, Sky June 20, 
S0-17, and S0-16 respectively.}
\end{center}
\end{figure}

\begin{figure}[tbp] \label{nwfig} \figurenum{6}
\begin{center}
\plotone{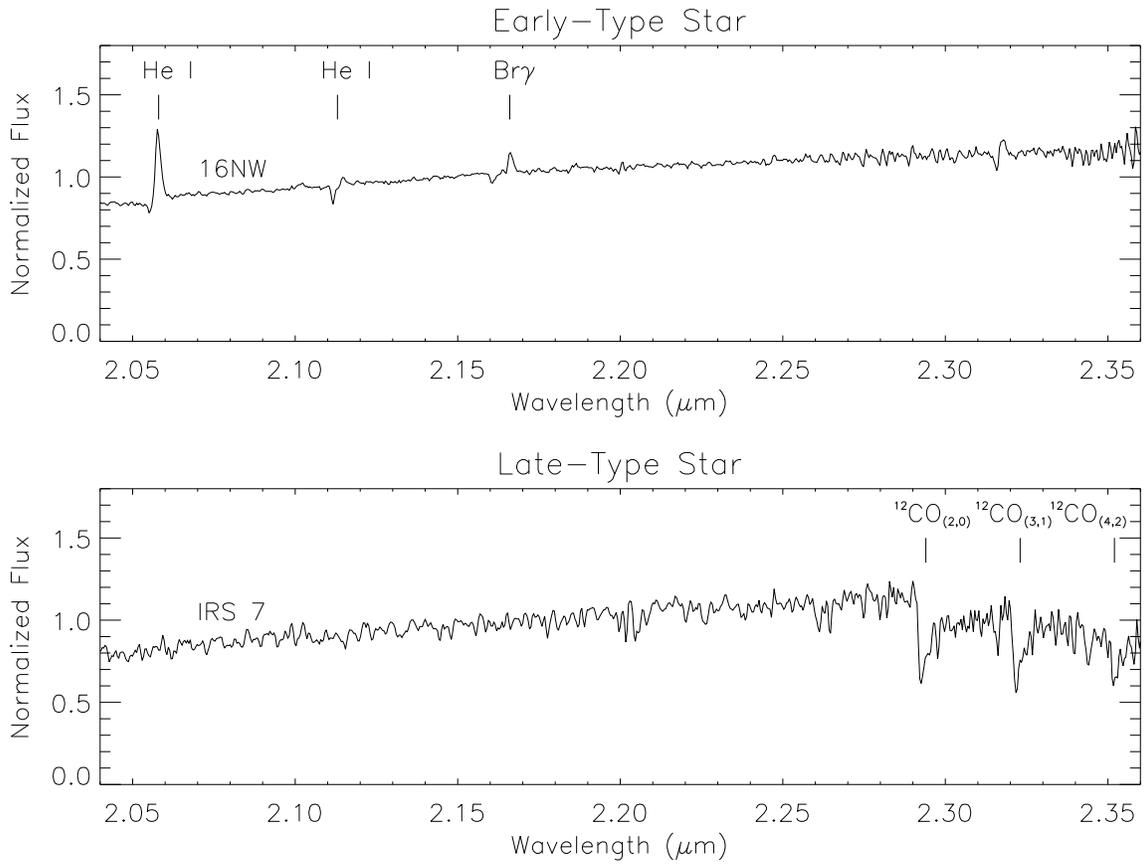}
\caption{Examples of bright early-type He I stars (IRS 16NW) and late-type 
supergiant stars 
(IRS 7) in the Galactic Center stellar population.}
\end{center}
\end{figure}

\begin{figure}[tbp] \label{corrfig} \figurenum{7}
\begin{center}
\plotone{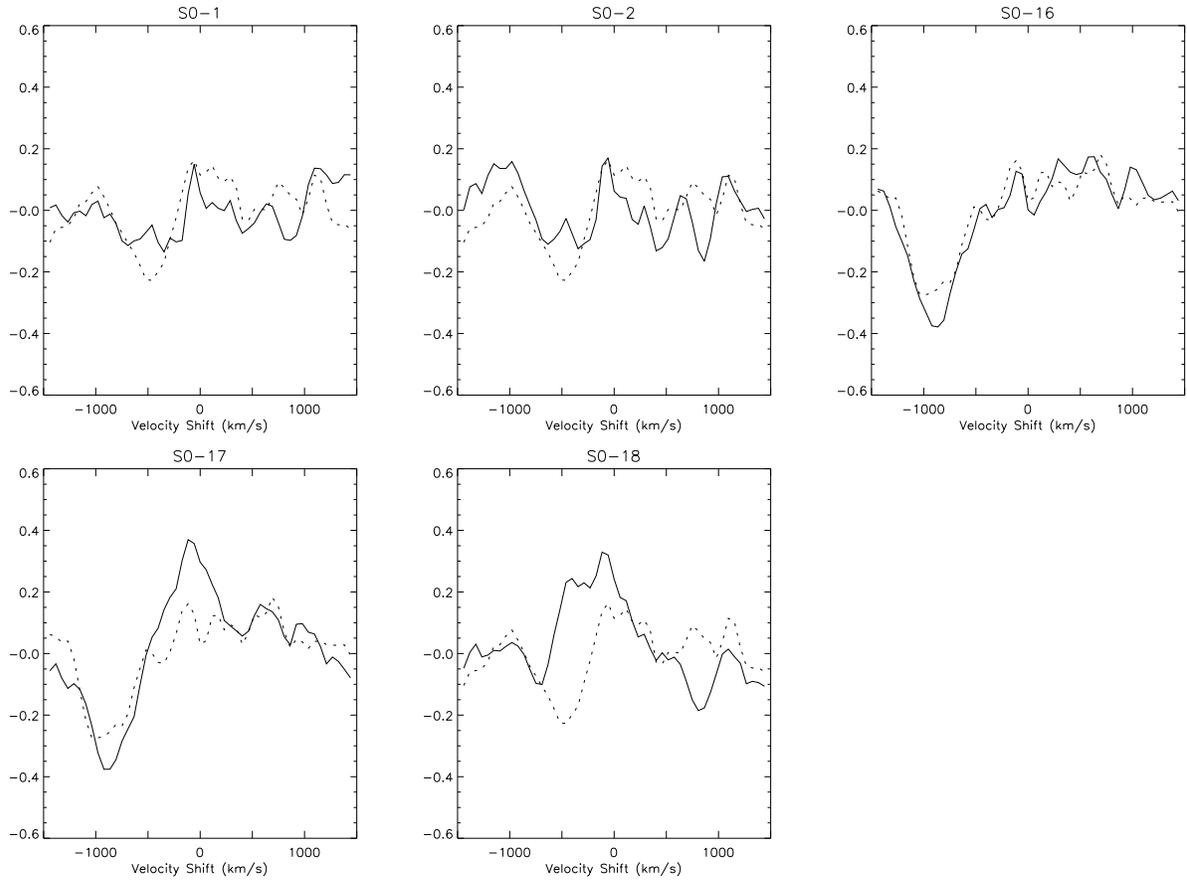}
\caption{Cross-correlation function of Sgr A* Cluster source spectra
with the K0-III standard spectrum from 2.29 to 2.37 \microns.  The
dotted line shows the cross-correlation function for the sky on the
same night of observations.}
\end{center}
\end{figure}

\begin{figure}[tbp] \label{asfig} \figurenum{8}
\begin{center}
\plotone{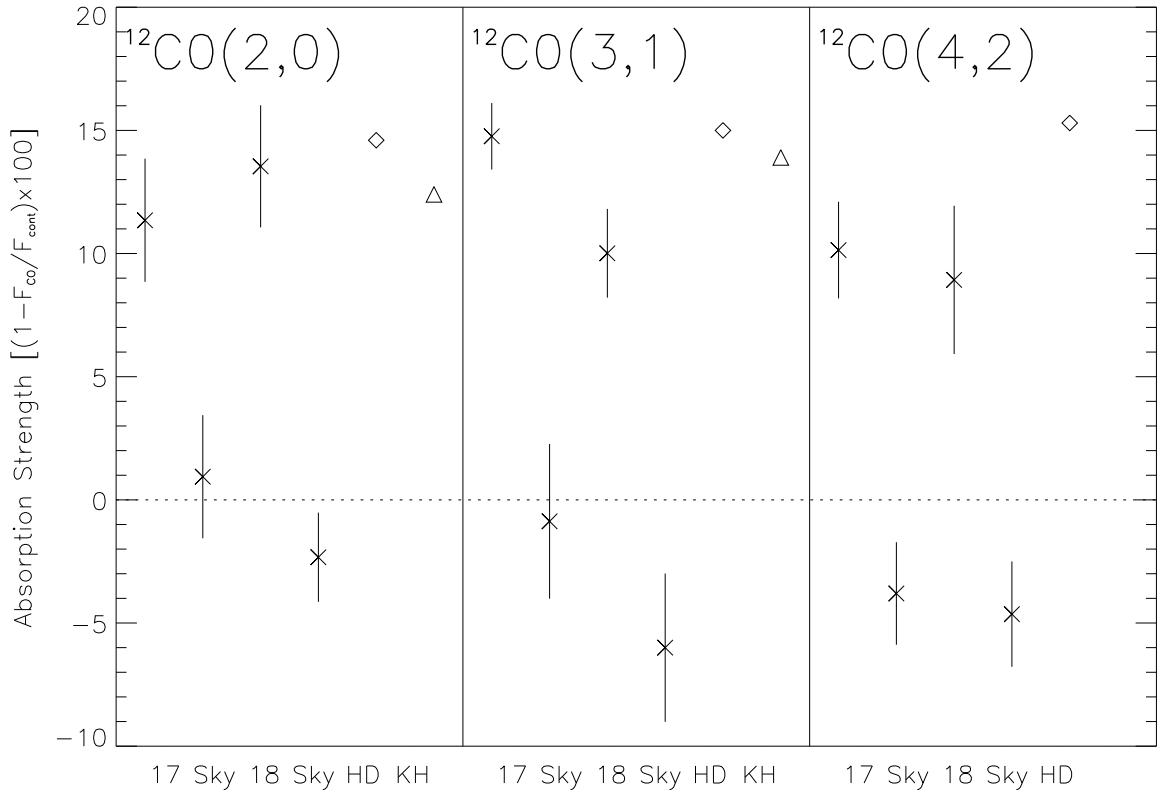}
\caption{CO Bandhead absorption measurements 
with 1 $\sigma$ error bars for S0-17 and S0-18 in comparison to 
the sky each night.  Diamonds
show measurements for the observed K0-III standard HD 203638 (1 $\sigma$
= 0.5), and 
triangles show the absorption strengths for K0-III standard {\it i} Cep 
(1 $\sigma$ = 0.1) from Kleinmann and Hall (1986).}
\end{center}
\end{figure}

\end{document}